\documentclass[final,3p,times,twocolumn]{elsarticle}

\usepackage{amssymb}
\usepackage{color}

\journal{Physics Letters B}

\begin{document}

\begin{frontmatter}



\title{Large Randall-Sundrum II Black Holes}

\author[label1,label2]{Shohreh Abdolrahimi}
\ead{abdolrah@ualberta.ca}
\author[label1,label3]{C\'{e}line Catto\"{e}n}
\ead{celine.cattoen-gilbert@canterbury.ac.nz}
\author[label1]{Don N. Page}
\ead{dpage@ualberta.ca}
\author[label1]{Shima Yaghoobpour-Tari}
\ead{yaghoobp@ualberta.ca}

\address[label1]{Department of Physics, 4-181 CCIS, University of Alberta,
Edmonton, Alberta T6G 2E1, Canada}
\address[label2]{ Institut f\"ur Physik, Universit\"at Oldenburg, Postfach 2503 D-26111 Oldenburg, 
Germany}
\address[label3]{BlueFern Supercomputing Unit, University of Canterbury, Christchurch 8140, New Zealand}


\author{}

\address{}

\begin{abstract}
Using a novel numerical spectral method, we have constructed an AdS$_5$-CFT$_4$ solution to the Einstein equation with a negative cosmological constant $\Lambda$ that is asymptotically conformal to the Schwarzschild metric. This method is independent of the Ricci-DeTurck-flow method used by Figueras, Lucietti, and Wiseman.  We have perturbed the solution to get large static black hole solutions to the Randall-Sundrum II (RSII) braneworld model.  Our solution agrees closely with that of Figueras et al. and also allows us to deduce the new results that to first order in $1/(-\Lambda M^2)$, the Hawking temperature and entropy of an RSII static black hole have the same values as the Schwarzschild metric with the same mass, but the horizon area is increased by about $4.7/(-\Lambda)$.

\end{abstract}

\begin{keyword}


\end{keyword}

\end{frontmatter}





\section{Introdcution}
The Randall-Sundrum II (RSII) braneworld model \cite{RSII} is one of the braneworld  models suggested for solving the hierarchy problem. The braneworld model includes a higher-dimensional spacetime that is called the bulk and a lower-dimensional spacetime called the brane, which is embedded in the bulk. All matter and fields in the standard model are supposed to propagate on the subspace manifold, the brane, but gravity is the only force that can propagate through the whole space, the bulk.  The RSII model is a warped five-dimensional braneworld model with an extra dimension that can be large. It is a very important question whether stationary large black holes exist within the RSII braneworld model.  If not, then the extremely strong observational evidence for very-nearly stationary astrophysical black holes would be a nearly conclusive reason for rejecting that theory as possibly physically realistic. After various conjectures and claims \cite{Tanaka,EFK,EGK,Yoshino,KKRS} that large black holes do not exist in the Randall-Sundrum II (RSII) braneworld model \cite{RSII}, Figueras and Wiseman \cite{FW} (henceforth FW) recently found such solutions by perturbing an AdS$_5$-CFT$_4$ solution that Figueras, Lucietti, and Wiseman \cite{FLW} (henceforth FLW) had found earlier by Ricci-DeTurck flow.  This AdS$_5$-CFT$_4$ metric is a solution to the Einstein equation with a negative cosmological constant $\Lambda$ that is asymptotically conformal to the Schwarzschild metric.  Because the Schwarzschild metric appears at an AdS$_5$ boundary with an infinite scale factor, it may be viewed as a black hole of infinite mass.

We had independently searched for and found the infinite-mass black hole solution by a different numerical method and were preparing to perturb it to get large-mass RSII black hole solutions when the Figueras et al.\ papers appeared. Here we report that our numerical solution agrees well with that of Figueras et al.\ and thus adds further evidence for the existence of large RSII black holes, despite the doubts expressed by previous work.

We used a spectral method, expressing the components of the 5-dimensional metric in terms of Legendre polynomials in the two nontrivial coordinates, with the appropriate boundary conditions imposed.  Then we chose the 210 coefficients of the polynomials to minimize the integrated square of the error of the Einstein equation, finding that we could reduce this by eight orders of magnitude from the case with no free parameters (constant polynomials). The integrated square is based on the Gauss-Legendre quadrature numerical method, and the minimization procedure uses the simplex search method for multivariable functions. This strongly suggests that we are numerically near an exact solution, though of course our limited computational resources meant that we could not use an infinite number of parameters to reduce the numerical error all the way to zero. This approach to solving Einstein equations is novel, and the good agreement of our results with the Figueras et al.\ results illustrates the success of the method, especially in comparison with the failure of various previous numerical attempts. 

We present an explicit approximate metric for the black hole on the brane. Using this approximate metric we demonstrate that the area of an RSII black hole on the brane is slightly greater than a black hole in pure four-dimensional general relativity, and  to leading order, the relations between the mass, Hawking temperature, and Bekenstein-Hawking entropy are precisely the same as in four-dimensional general relativity.  In other words, astrophysical-sized black holes in the Randall-Sundrum II braneworld model are extremely close to what four-dimensional general relativity would predict.  Although this may be disappointing for those hoping to distinguish between the two models by observations of black holes, it it highly encouraging for those who have postulated that the Randall-Sundrum II braneworld model is a model consistent with our observations.

\section{ Infinite Black Hole Metric}
For brevity of notation, we use units in which the 5-dimensional bulk cosmological constant is $\Lambda = -6$.  We start with the AdS-black string metric \cite{CHR}
\begin{eqnarray}
ds^2 &=& \frac{1}{w^2}~~\big[dw^2 + U(r)^{-1}dr^2 - U(r)dt^2 \nonumber \\&+& r^2d\Omega^2 \big],
\end{eqnarray}
where $U(r) = 1 - 2M/r$ and where $d\Omega^2 = d\theta^2 + \sin^2{\theta}d\phi^2$ is the unit two-sphere metric.  Letting $r = 2M/y$ and $w = 2M/v$ gives
\begin{eqnarray}
ds^2&=&\frac{dv^2}{v^2}+\frac{v^2 dy^2}{y^4(1-y)}\nonumber \\
&-&4v^2(1-y)dt^2 +\frac{v^2}{y^2}d\Omega^2.
\end{eqnarray}

The hypersurfaces of constant $v$ are Schwarzschild metrics of mass $m(v) = v/2$.  The curvature at $y > 0$ diverges at $v=0$, so this black string metric is singular.  We modify the metric by adding some $y^2$ terms to remove this singularity, and we also introduce four metric functions to give
\begin{eqnarray}
ds^2 &=& A\frac{dv^2}{v^2 + y^2} +B\frac{(v^2+y^2)dy^2}{y^4(1-y)} \nonumber \\ &-&4C(v^2+y^2)(1-y)dt^2 +D\frac{v^2}{y^2}d\Omega ^2.
\end{eqnarray}
We then replace $v$, which ranges from 0 to $\infty$, by $x = y^2/(y^2 + v^2)$, so that the metric becomes
\begin{eqnarray}
ds^2 &=&A(1-x){\left[\frac{dx}{2x(1-x)}-\frac{dy}{y}\right]}^2\nonumber \\
         &+& B\frac{dy^2}{xy^2(1-y)}-4C\frac{y^2(1-y)}{x}dt^2\nonumber \\
         &+& D\frac{1-x}{x}d\Omega ^2,
\end{eqnarray}
where $0\leq x\leq1$, $0\leq y\leq1$ and $A$, $B$, $C$ and $D$ are smooth functions of $x$ and $y$.  The coordinate boundaries are these:  $x=0$ is the infinite AdS boundary that is conformal to the Schwarzschild metric when we impose $A = B = C = D = 1$ there, $y=0$ is the extremal Poincare horizon, $x=1$ is the axis of symmetry where the two-sphere shrinks to zero size and where we impose $A = D$ for regularity, and $y=1$ is the black hole horizon where we impose the regularity requirement $B = C$.

The most general metric satisfying all the symmetries for our problem has five components. On the other hand, since the metric functions depend non-trivially on the two coordinates, $x$ and $y$, and the choice of these is gauge dependent, one can reduce the number of the metric components to three. The common method for finding a unique solution for the Einstein equation numerically is to fix the gauge before discretization. Otherwise, one will have a family of solutions parametrized by one function that is gauge dependent. But in our case, we assume four unknown functions instead of three as mentioned, and we still get a unique solution. Our explanation for this result can be related to our restriction of $A(x,y)$, $B(x,y)$, $C(x,y)$, and $D(x,y)$ to polynomials for simplicity; we also tried rational functions, but they did not seem to work numerically so well. Having polynomials of some fixed finite order means that with such restricted functions, some gauges are better than others. For clarification, we can consider the case of a spherically symmetric static metric
\begin{equation}
ds^2=-A(x)dt^2+\frac{1}{B(x)}dx^2+\frac{1}{C(x)}d\Omega ^2.
\label{eqr1}
\end{equation} 
If we consider the restriction for $A(x,y)$, $B(x,y)$, and $C(x,y)$ to be polynomials, then with having all three functions, one can find $A(x)=1-x$, $B(x)=x^4-x^5$, and $C(x)=x^2$ solves the vacuum Einstein equations. However, if one choose the gauge $B(x)=1$, no polynomials of finite order would give an exact solution, and we would expect greater error. On the other hand, we are not looking for an exact solution, so with our restriction to have a fixed order of polynomials for each function, surely we would get a better result with more functions, even if for an exact solution one or more functions would be just gauge.

We impose these regularity conditions and also solve the Einstein equation to lowest order in $x$ by writing
\begin{eqnarray}
A\!&=&\!1\!-\!x(1\!-\!x)(1\!+\!2f(y))\!+\!x^2g(y)\!\nonumber \\
&+&\!x^2(1\!-\!x)\tilde{A}(x,y),
\nonumber \\ 
B\!&=&\!1\!+\!x f(y)\!+\!x^2\tilde{B}(x,y), \nonumber \\
C\!&=&\!1\!+\!x f(y)\!+\!x^2\tilde{B}(x,y)\!+\!x^2(1\!-\!y)\tilde{C}(x,y),
\nonumber \\
D\!&=&\!1\!+\!x(1\!-\!x)(1\!+\!f(y))\!+\!x^2g(y)\! \nonumber \\
&+&\!x^2(1\!-\!x)\tilde{D}(x,y).
\end{eqnarray}

With units such that $\Lambda = -6$, the vacuum Einstein equation in the 5-dimensional bulk is
\begin{equation}
E_{\alpha\beta} \equiv R_{\alpha \beta}+4g_{\alpha \beta} = 0.
\end{equation}
We define the integrated square error of the Einstein equation to be
\begin{equation}
I = \int {E_{\alpha\beta} E^{\alpha\beta} \sqrt{-^{(5)}g}\, d^5x},
\end{equation}
where we choose $\Delta t = 2\pi$ in order to get a definite finite integral (assuming that $E_{\alpha\beta} E^{\alpha\beta}$ falls off fast enough toward the infinite AdS boundary at $x=0$, where the metric determinant $^{(5)}g \propto 1/x^6$ diverges, so that the integral converges).

We choose polynomials for the functions $f(y)$, $g(y)$, $\tilde{A}(x,y)$, $\tilde{B}(x,y)$, $\tilde{C}(x,y)$, and $\tilde{D}(x,y)$ and numerically vary the coefficients to minimize the integrated square error $I$.  For $A=B=C=D=1$, $I \approx 4038$, but when we went up to sixth-order polynomials with a total of 210 coefficients, the integrated squared error was reduced to $0.00004238$, nearly eight orders of magnitude smaller.  The maximum value of the squared error at any point within the 5-dimensional spacetime was then $E_{\alpha\beta} E^{\alpha\beta} = 0.000154$.  Thus we appear to have strong evidence that our numerical method is converging toward an exact solution of the infinite black hole metric.

Because our metric uses different coordinates from those used by FLW, it is not easy to make many comparisons over the entire bulk 5-dimensional manifold.  We have found that the minimum value for the length scale given by the inverse fourth root of the total trace of the square of the Weyl tensor, $(C_{\alpha\beta\gamma\delta}C^{\alpha\beta\gamma\delta})^{-1/4}$, is about 0.206 in our metric, which is within 4\% of the value 0.198 that FLW privately reported to us from their metric.  However, we shall make many more comparisons below for the 4-dimensional large black hole metric.

\section{Finite Black Hole Metrics}
To get a black hole metric with a large but finite mass, we need to replace the infinite AdS boundary at $x=0$, where the metric is conformally Schwarzschild but with infinite mass, by an RSII brane with induced metric $\gamma_{\mu\nu}$, and with a mirror image of the bulk metric on the opposite side of the brane.  Assuming no matter on the brane, in our units with $\Lambda = -6$, the Israel junction condition is $[K_{\mu\nu}] = - 2\gamma_{\mu\nu}$, where the square brackets denote the difference in the second fundamental form $K_{\mu\nu}$ from the back side to the front side, where it has the opposite value.  Hence $K_{\mu\nu} = -\gamma_{\mu\nu}$ on the front side, where we are using the opposite sign convention from FW.

We write the asymptotic form of the bulk metric near the infinite AdS boundary as
\begin{eqnarray}
ds^2 = \frac{1}{z^2}\big[dz^2 + \tilde{g}_{\mu\nu}(z,x)dx^\mu dx^\nu\big],
\end{eqnarray}
where the $z$ is the exponential of the negative of the outward proper distance as one approaches the AdS boundary at infinite proper distance (where $z \rightarrow 0$), and where $x$ denotes the other four coordinates (not just the single $x$ coordinate used above).  Then we use just enough of the Fefferman-Graham (FG) expansion \cite{FG,Graham,HSS,FW} to write
\begin{eqnarray}
\tilde{g}_{\mu\nu}(z,x) &\approx & g_{\mu\nu}^{(0)}(x)
 + z^2\big[-\frac{1}{2}R_{\mu\nu}^{(0)}(x) \nonumber \\ 
 &+& \frac{1}{12}R^{(0)}g_{\mu\nu}^{(0)}(x)\big]+z^4 t_{\mu\nu}(x).
\end{eqnarray}
Here $g_{\mu\nu}^{(0)}(x)$ is the conformal metric on the infinite AdS boundary, and $t_{\mu\nu}(x)$ is a tensor that is covariantly conserved in the conformal metric $g_{\mu\nu}^{(0)}(x)$ and which gives the second set of constants of integration for the bulk Einstein equation when written as second-order equations in $z$.  We extracted polynomial forms for $t_{\mu\nu}(x)$ at the boundary from our sixth-order polynomial metric coefficients for the bulk metric.

The infinite-mass black hole bulk solution is equivalent to putting the RSII brane at $z=0$ and setting $g_{\mu\nu}^{(0)}(x) = g_{\mu\nu}^\mathrm{Sch}$, the Schwarzschild metric with unit horizon radius.  Since the Schwarzschild metric is Ricci-flat, $R_{\mu\nu}^{(0)}(x)$ vanishes, so then there is no $z^2$ term in the FG expansion.  One can also show that then $t_{\mu\nu}(x)$ is traceless as well as conserved \cite{FG,Graham,HSS,FW}.  An approximation for it can be extracted from the numerical data for the bulk solution found above.

For a large but finite mass black hole, we put the brane at small $z = \epsilon$ and perturb the conformal metric at $z=0$ to $g_{\mu\nu}^{(0)} = g_{\mu\nu}^\mathrm{Sch} + \epsilon^2 h_{\mu\nu}$.  This gives the Ricci tensor a small perturbation, so that now it is of order $\epsilon^2$, and the $z^2$ term in the FG expansion is no longer zero.  The perturbation in $g_{\mu\nu}^{(0)}$ also perturbs $t_{\mu\nu}(x)$ away from its traceless form, but since that term is multiplied by $z^4$ in the FG expansion, for results to lowest nontrivial order in $\epsilon$, it is sufficient to use the original value of $t_{\mu\nu}(x)$ from the infinite-mass black hole bulk solution.  

Using the fact that $K_{\mu\nu} = (z/2)\partial_z[\tilde{g}_{\mu\nu}(z,x)/z^2]$, to lowest nontrivial order in $\epsilon$ the Israel junction condition implies that $R_{\mu\nu}^{(0)}(x) - (1/6)R^{(0)}g_{\mu\nu}^{(0)}(x) \approx 4\epsilon^2 t_{\mu\nu}(x)$.  Since to this order $t_{\mu\nu}(x)$ is traceless, one further gets that the Ricci tensor of the perturbed conformal metric $g_{\mu\nu}^{(0)}(x) = g_{\mu\nu}^\mathrm{Sch} + \epsilon^2 h_{\mu\nu}$ is $R_{\mu\nu}^{(0)}(x) \approx 4\epsilon^2 t_{\mu\nu}(x)$, which with a knowledge of $t_{\mu\nu}(x)$ is sufficient to determine $h_{\mu\nu}$ and hence the spherically symmetric static metric $g_{\mu\nu}^{(0)}(x)$.  Then the induced metric on the brane is
\begin{equation}
\gamma_{\mu\nu} = \frac{1}{\epsilon^2}\tilde{g}_{\mu\nu}
= \frac{1}{\epsilon^2}g_{\mu\nu}^\mathrm{Sch} + h_{\mu\nu}
+ O(\epsilon^2).
\end{equation}

The bulk Einstein equation plus the Israel junction condition for a brane with the RSII value of the tension and without matter imply that the Ricci scalar of the brane metric is zero.  We can achieve this to first order in the perturbation for a generic static spherically symmetric metric on the brane by going to the gauge $h_t^t = 0$ (which is equivalent to choosing the coordinate $y$ so that on the brane $g_{tt} = -(1-y)$ after rescaling $t$ so that $g_{tt} = -1$ at radial infinity, $y=0$), defining $h_\theta^\theta = h_\phi^\phi = (y^2/6)F(y)$, and then setting
\begin{eqnarray}
h_y^y &=& -\frac{2y^2(1-y)}{3(4-3y)}\left(F+y\frac{dF}{dy}\right).
\end{eqnarray}

If we now define $(2M)^2 \equiv 1/\epsilon^2 = 6/(-\Lambda\epsilon^2)$ after reverting to general units in this last expression, and if we define a new radial coordinate $\rho = 2M/y$, then to first order in $\epsilon^2 = (3/2)/(-\Lambda M^2)$, we can write the metric on the brane (after rescaling the time coordinate $t$ by a factor of $4M$) as
\begin{eqnarray}
&&^4ds^2 = \gamma_{\mu\nu}dx^\mu dx^\nu \nonumber \\
        &=&\!\! \left[1\! -\! \frac{1}{(\!-\Lambda\rho^2)}
                     \frac{\rho\!-\!2M}{\rho\!-\!1.5M}\!
		     \left(\!F\!-\rho\frac{dF}{d\rho}\!\right)\!\right]
		     \!\left(\!1\! -\! \frac{2M}{\rho}\!\right)^{\!-1}\!d\rho^2 		        \nonumber \\
	&-& \left(1 - \frac{2M}{\rho}\right) dt^2
          + \left[\rho^2 + \frac{1}{(-\Lambda)}F\right]d\Omega^2.
\end{eqnarray}

One can show that the asymptotic behaviour of $t_{\mu\nu}(x)$ (which goes as $1/\rho^5$ for $\rho \gg 2M$ with known coefficients \cite{FLW,FW}) implies that $F$ approaches unity as $\rho \rightarrow \infty$ or $y \rightarrow 0$.  To fit the FLW numerical data $t_{\mu\nu}^{(1)}(x) = t_{\mu\nu}^{\mathrm{FLW}}(x)$ which they kindly sent us, and to fit our numerical data $t_{\mu\nu}^{(2)}(x) = t_{\mu\nu}^{\mathrm{our}}(x)$, we took $F_1 = F_{\mathrm{our}}$ and $F_2 = F_{\mathrm{FLW}}$ to be cubic polynomials in $y \equiv 2M/\rho$ with the constant coefficient set to unity and then chose the other three coefficients in each case to minimize the respective
\begin{equation}
J_i ~~=~~\frac{ \int\! {\rho^4 \Delta t_{\mu\nu}^{(i)} \Delta t^{\mu\nu}_{(i)} \sqrt{-^{(4)}\gamma} d^4x}}{ \int\!\! {\rho^4 t_{\mu\nu}^{\mathrm{FLW}} t^{\mu\nu}_{\mathrm{FLW}}      
            \sqrt{-^{(4)}\gamma} d^4x}},
\label{J}
\end{equation}
where for each of the two values of $i$ ($i=1$ for the FLW data and $i=2$ for our data) in the numerator $\Delta t_{\mu\nu}^{(i)} = t_{\mu\nu}^{F_i} - t_{\mu\nu}^{(i)}$ is the difference between the $t_{\mu\nu}^{F_i}(x)$ given by the cubic for $F_i(x)$ and the $t_{\mu\nu}^{(i)}$ given by the numerical data. The integral in the denominator was included to make $J$ a normalized mean-square error and has $t_{\mu\nu}^{\mathrm{FLW}}(x)$ given by the FLW numerical data, the same in each case to give a constant normalizing factor.  The factor of $\rho^4$ was included to increase the weight of the large-$\rho$ part, though the integrals are still dominated by the small-$\rho$ part, since $t_{\mu\nu}(x)$ drops off asymptotically as the inverse fifth power of the radial coordinate $\rho$ \cite{FLW,FW}.

For the FLW numerical data $t_{\mu\nu}^{\mathrm{FLW}}(x)$ (which was constrained to be traceless and proved to be very nearly conserved and matching the predicted $y^5$ dependence at small $y$), $J_1$ was minimized at $J_{\mathrm{FLW}} \approx 0.0000620$ for
\begin{eqnarray}
F_{\mathrm{FLW}}\! &\approx &\! 1\! -\! 1.062 \left(\frac{2M}{\rho}\right)
 \!+\! 0.554 \left(\frac{2M}{\rho}\right)^2 \nonumber \\
 \! &-& \! 0.120 \left(\frac{2M}{\rho}\right)^3\!\!.
\end{eqnarray}
For our numerical data $t_{\mu\nu}^{\mathrm{our}}(x)$ (which was not quite traceless and conserved and also had a small spurious $y^4$ term), the normalized mean-square error $J_2$ was minimized at $J_{\mathrm{our}} \approx 0.00139$ for
\begin{eqnarray}
F_{\mathrm{our}}\! &\approx &\! 1\! -\! 1.002\left(\frac{2M}{\rho}\right)
 \!+\! 0.434 \left(\frac{2M}{\rho}\right)^2 \nonumber \\
 \! &-& \! 0.059 \left(\frac{2M}{\rho}\right)^3\!\!.
\end{eqnarray}
\begin{figure}[htb]
\begin{center}
\includegraphics[width=7cm]{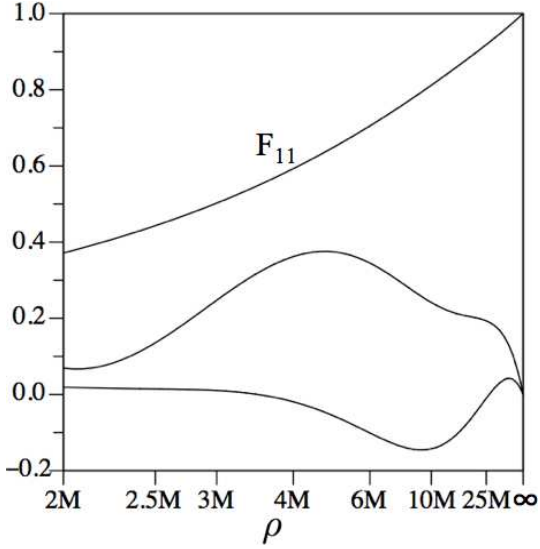}
\caption{At the top is an 11th-order polynomial fit $F_{11}$ to the FLW data, which gave normalized mean-square error $J_{11} = 0.0000572$, 92\% of $J_{\mathrm{FLW}}$.  Because the differences from $F_{11}$ of the cubic fits $F_{\mathrm{FLW}}$ and $F_{\mathrm{our}}$ are too small to show up when plotted directly on this graph, at the bottom we have expanded these differences by a factor of 50 and plotted $50(F_{\mathrm{FLW}} - F_{11})$ (bottom curve) and $50(F_{\mathrm{our}} - F_{11})$ (middle curve).}\label{F5} 
\end{center} 
\end{figure} 
From the fact that $J_{\mathrm{our}} \approx 22\, J_{\mathrm{FLW}}$, clearly our data is less accurate than the FLW data, which is not surprising since we varied only 210 parameters in our spectral method, whereas FLW used grids of $40\times 40$ (or 1\,600 points) and of $160\times 160$ (or 25\,600 points).  Also, the individual coefficients of these two cubics have large relative differences, but the ratio of the two cubics themselves never differs by more than 1.3\% from unity, so they show good agreement between what is generated by our numerical data and by what is given by the FLW data.

We also used the integral of Eq.\ (\ref{J}) with $\Delta t_{\mu\nu}^{(3)} = t_{\mu\nu}^{F_{\mathrm{our}}} - t_{\mu\nu}^{\mathrm{FLW}}(x)$, the difference between the stress tensor $t_{\mu\nu}^{F_{\mathrm{our}}}$ generated by our $F_{\mathrm{our}}$ fit to our data and the stress tensor $t_{\mu\nu}^{\mathrm{FLW}}(x)$ given directly by the FLW data.  This gave $J_3 = J_{\mathrm{our\ fit\ vs.\ FLW\ data}} \approx 0.000214 \approx 3.4\, J_{\mathrm{FLW}}$, so the $t_{\mu\nu}^{F_{\mathrm{our}}}(x)$ generated by our $F_{\mathrm{our}}(x)$ fits the FLW data about 6.5 times better than it fits our data.  This is also not surprising, since the $t_{\mu\nu}^{F_{\mathrm{our}}}(x)$ generated by our $F_{\mathrm{our}}(x)$ fit was constrained to be both traceless and conserved, whereas the $t_{\mu\nu}^{\mathrm{our}}(x)$ extracted directly from our data was not.  

Furthermore, we calculated the mean-square error between the $t_{\mu\nu}^{F_{\mathrm{our}}}(x)$ generated by our $F_{\mathrm{our}}$ fit to our data and the $t_{\mu\nu}^{F_{\mathrm{FLW}}}(x)$ generated by the $F_{\mathrm{FLW}}$ fit to the FLW data, using in Eq.\ (\ref{J}) $\Delta t_{\mu\nu}^{(4)} = t_{\mu\nu}^{F_{\mathrm{our}}} - t_{\mu\nu}^{F_{\mathrm{FLW}}}(x)$, and got $J_4 = J_{\mathrm{our\ fit\ vs.\ FLW\ fit}} \approx 0.000146 \approx 2.4\, J_{\mathrm{FLW}}$, so the $t_{\mu\nu}^{F_{\mathrm{our}}}(x)$ generated by our $F_{\mathrm{our}}$ fits the $t_{\mu\nu}^{F_{\mathrm{FLW}}}(x)$ generated by the $F_{\mathrm{FLW}}$ fit to the FLW data nearly 9 times better than it fits the $t_{\mu\nu}^{\mathrm{our}}(x)$ directly extracted from our data, which is not quite traceless and conserved, as the $t_{\mu\nu}^{F_{\mathrm{our}}}(x)$ generated by the fitting $F_{\mathrm{our}}$ is constrained to be.

We also calculated the ratios between the values of each of the three individual components of $t_{\mu\nu}^{\mathrm{FLW}}(x)$ as given by the FLW data, the fit to the FLW data given by $t_{\mu\nu}^{F_{\mathrm{FLW}}}(x)$, and the fit to our data given by $t_{\mu\nu}^{F_{\mathrm{our}}}(x)$.  These ratios were generally within 1-2\% of unity, with the maximum differing by less than 2.9\%.  The ratio of the $h_y^y$'s generated by $F_{\mathrm{our}}$ and by $F_{\mathrm{FLW}}$, which involves a derivative of $F$ as given in Eq.\ $(11)$, did differ by up to about 9.3\%, so when one is neither near the black hole horizon ($\rho = 2M$ exactly in our gauge) nor near $2M/\rho = 0$ (at both of which limits $h_y^y = 0$), the deviation of $g_{\rho\rho}$ from the Schwarzschild value may not be given to very high precision by our $F_{\mathrm{our}}$.  However, the other results appear to agree within a very few percent from those given by FLW, which gives strong independent confirmation of their results.

One can readily calculate from the metric $(12)$ for any $F$ that has only a constant term and negative powers of $\rho$, as $F_{\mathrm{FLW}}$ given by Eq.\ $(14)$ and $F_{\mathrm{our}}$ given by Eq.\ $(15)$ do, that the ADM mass is precisely $M$ and that the surface gravity of the black hole horizon is exactly $1/(4M)$, the same as for the Schwarzschild metric.  Of course, aside from the numerical approximations for determining $F$, the metric $(12)$ is only correct to first order in our perturbation parameter $1/(-\Lambda M^2)$, so there might be corrections to the surface gravity of a static RSII black hole to second order in $1/(-\Lambda M^2)$.  However, one can deduce that to first order in $1/(-\Lambda M^2)$, the Hawking temperature and entropy for the RSII black hole have the same values as they do for the Schwarzschild metric.

On the other hand, the horizon area is shifted from the Schwarzschild value $A_\mathrm{Sch} = 4\pi(2M)^2$ to $A_\mathrm{RSII} = 4\pi[(2M)^2 + F(1)/(-\Lambda)]$, where $F(1)$ is the value of $F$ on the horizon, at $y \equiv 2M/\rho = 1$.  The fit to the FLW numerical data gives $F_{\mathrm{FLW}}(1) \approx 0.372$, and the fit to our numerical data gives $F_{\mathrm{our}}(1) \approx 0.373$, which agree within about 0.3\%.  Therefore, the change from a Schwarzschild black hole to an RSII black hole on a brane with the same ADM mass $M$ increases the horizon area (but not the Hawking entropy) by the amount $\Delta A = 4\pi F(1)/(-\Lambda) \approx 4.67/(-\Lambda)$, where here we used the FLW data value as probably more accurate.  (The value from our data would give a coefficient of about 4.69 rather than 4.67.)

\section{Conclusion}
We have provided independent numerical evidence in support of the numerical discovery of Figueras and Wiseman \cite{FW} of the existence of large static black holes in the Randall-Sundrum II braneworld model \cite{RSII}, by a significantly different numerical method.  Our results agree quite well with theirs, such as giving an increase in the black hole horizon area over that of the Schwarzschild metric of the same mass by $4.69/(-\Lambda)$ or $4.67/(-\Lambda)$ respectively.  (This is a new result, not reported in  \cite{FW}.)  We have obtained a good closed-form approximation to the metric of the black hole on the brane, Eqs.\ $(12)$ and either $(14)$ or $(15)$.  We have also shown the new result that to first order in our perturbation parameter $1/(-\Lambda M^2)$, the Hawking temperature and entropy of the black hole is the same as that of a Schwarzschild black hole of the same ADM mass $M$.

If large black holes did not exist in the Randall-Sundrum II braneworld model, the astrophysical observations of such black holes would have been strong evidence against the viability of that model.  However, our confirmation of the large black holes in RSII found by Figueras and Wiseman \cite{FW}, and the fact that they are very nearly the same as Schwarzschild black holes, show that the RSII model is not excluded in this respect.

We have benefited from conversations with Pau Figueras, James Lucietti, and Toby Wiseman and greatly appreciate their sharing their detailed numerical data with us for comparison.  CC acknowledges an Avadh Bhatia Postdoctoral Fellowship at the University of Alberta.  This research was also supported in part by the Natural Sciences and Engineering Research Council of Canada.



\bibliographystyle{elsarticle-num}
\bibliography{<your-bib-database>}

\begin{thebibliography}{00}

\bibitem{RSII} L. Randall and R. Sundrum, Phys.\ Rev.\ Lett.\ {\bf 83}, 4690 (1999).

\bibitem{Tanaka} T. Tanaka, Prog.\ Theor.\ Phys.\ Suppl.\ {\bf 148}, 307 (2002).

\bibitem{EFK} R. Emparan, A. Fabbari, and N. Kaloper, J. High Energy Phys.\ {\bf 08}, 043 (2002).

\bibitem{EGK} R. Emparan, J. Garcia-Bellido, and N. Kaloper, J. High Energy Phys.\ {\bf 01}, 079 (2003).

\bibitem{Yoshino} H. Yoshino, J. High Energy Phys.\ {\bf 01}, 068 (2009).

\bibitem{KKRS} B. Kleihaus, J. Kunz, E. Radu, and D. Senkbeil, Phys.\ Rev.\ D {\bf 83}, 104050 (2011).


\bibitem{FW} P. Figuras and T. Wiseman, Phys.\ Rev.\ Lett.\ {\bf 107}, 081101 (2011).

\bibitem{FLW} P. Figuras, J. Lucietti, and T. Wiseman, Class.\ Quantum Grav.\ {\bf 28}, 215018 (2011).

\bibitem{CHR} A. Chamblin, S. W. Hawking, and H. S. Reall, Phys.\ Rev.\ D {\bf 61}, 065007 (2000).

\bibitem{FG} C. Fefferman and C, R. Graham, ``Conformal Invariants,'' in {\it Elie Cartan et les Math'{e}matiques d'Aujourd'hui}, Ast'{e}risque, hors s'{e}rie, p.\ 95 (1985).

\bibitem{Graham} C. R. Graham, ``Volume and Area Renormalizations for Conformally Compact Einstein Metrics,'' 19th Winter School on Geometry and Physics, Srni, Czech Republic, 1999, arXiv:math/9909042[math-dg].

\bibitem{HSS} S. de Haro, K. Skenderis, and S. N. Solodukhin, Commun.\ Math.\ Phys.\ {\bf 217}, 595-622 (2001).

 \end{thebibliography}



\end{document}